\documentclass[aps,prl,twocolumn,showpacs,groupedaddress]{revtex4}  % for review
\usepackage{graphicx}  % needed for figures
\usepackage{dcolumn}   % needed for some tables
\usepackage{bm}        % for math
\usepackage{amssymb}   % for math

\newcommand{\simleq}{\; \raisebox{-0.4ex}{\tiny$\stackrel{{\textstyle<}}{\sim}$}\;}
\def\MET{{\mbox{$E\kern-0.57em\raise0.19ex\hbox{/}_{T}~$}}}
\def\METnoSpace{{\mbox{$E\kern-0.57em\raise0.19ex\hbox{/}_{T}$}}}

\begin{document}

%\setpagewiselinenumbers
%\modulolinenumbers[1]
%\linenumbers

% The following information is for internal review, please remove them for submission
%\leftline{Version 1.0 as of \today} 
%\leftline{Primary authors: Yuri Gershtein, Yurii Maravin}
%\leftline{To be submitted to PRL}
%\rightline{Comment to {\tt d0-run2eb-013@fnal.gov}}
%\rightline{by May 30, 2008}

% the following line is for submission, including submission to the arXiv!!
\hspace{5.2in} \mbox{Fermilab-Pub-08-169-E}

\title{Search for long-lived particles decaying into electron or photon pairs with the D0 detector}
% LIST_OF_AUTHORS_R2.TEX               5/23/08              
%
\author{V.M.~Abazov$^{36}$}
\author{B.~Abbott$^{75}$}
\author{M.~Abolins$^{65}$}
\author{B.S.~Acharya$^{29}$}
\author{M.~Adams$^{51}$}
\author{T.~Adams$^{49}$}
\author{E.~Aguilo$^{6}$}
\author{M.~Ahsan$^{59}$}
\author{G.D.~Alexeev$^{36}$}
\author{G.~Alkhazov$^{40}$}
% ...

\affiliation{\vspace{0.1 in}(The D\O\ Collaboration)\vspace{0.1 in}}
% ...
\affiliation{$^{6}$University of Alberta, Edmonton, Alberta, Canada,
                Simon Fraser University, Burnaby, British Columbia, Canada,
                York University, Toronto, Ontario, Canada, and
                McGill University, Montreal, Quebec, Canada}
% ...
\affiliation{$^{29}$Tata Institute of Fundamental Research, Mumbai, India}
% ...
\affiliation{$^{36}$Joint Institute for Nuclear Research, Dubna, Russia}
% ...
\affiliation{$^{40}$Petersburg Nuclear Physics Institute,
                St. Petersburg, Russia}
% ...
\affiliation{$^{49}$Florida State University, Tallahassee, Florida 32306, USA}
% ...
\affiliation{$^{51}$University of Illinois at Chicago,
                Chicago, Illinois 60607, USA}
% ...
\affiliation{$^{59}$Kansas State University, Manhattan, Kansas 66506, USA}
% ...
\affiliation{$^{65}$Michigan State University,
                East Lansing, Michigan 48824, USA}
% ...
\affiliation{$^{75}$University of Oklahoma, Norman, Oklahoma 73019, USA}
% ...
  % input Dzero author list
\date{\today}

\begin{abstract}
In this Letter we report on a search for long-lived particles 
that decay into final states with two electrons or photons. 
Such long-lived particles arise in a variety of theoretical models, like 
hidden valleys and supersymmetry with  gauge-mediated breaking.  
By precisely reconstructing the direction of the electromagnetic 
shower we are able to probe much longer lifetimes than previously 
explored. We see no evidence of the existence 
of such long-lived particles and interpret 
this search as a quasi model-independent limit on their production 
cross section, as well as a limit
on a long-lived fourth generation quark.

\end{abstract}

\pacs{13.85.Qk,13.85.Rm,14.65.-q,14.80.-j}
\maketitle 

%\section{\label{sec:level1}First-level heading}

The standard model is surprisingly successful in describing
phenomena observed at accelerators. One would expect, given its numerous 
theoretical shortcomings and the proliferation of searches for deviations 
from it, that a more general underlying theory would have been already 
revealed. It is therefore a possibility that the discovery of new physics 
eludes us because the new physics looks different from popular standard 
model extensions like minimal supersymmetry (SUSY).

In this Letter we search for pairs of electromagnetic (EM) showers 
from electrons or photons that originate from the same point in
space, away from the $p\bar{p}$ interaction point. Such events 
can be a signature of a long-lived $b^\prime$ quark decaying into a $Z$ boson 
and a jet~\cite{bprime_theory}. In models with gauge-mediated SUSY 
breaking~\cite{gmsb_theory} a long-lived neutralino with large 
higgsino component can decay into a $Z$ boson and a gravitino. 
In the hidden valley models~\cite{hv_theory}, $v$-mesons can decay 
into electron pairs. In all of the above examples, a significant 
imbalance in transverse energy can be present due to 
$Z$ boson or $v$ hadron decays into neutrinos or lightest 
supersymmetric particles (LSP) that remain undetected. 

A search for such long-lived particles at hadron colliders was performed 
by CDF ~\cite{cdf_bprime} based on the reconstruction of lepton 
tracks from a secondary vertex. The sensitivity to large lifetimes in that 
search is limited by the difficulties in reconstructing tracks that originate 
far from the interaction point. In our analysis, we use the
fine segmentation of the D0 detector  to reconstruct the directions 
of the EM showers and use that to reconstruct the common vertex. This method 
allows us to probe dramatically longer decay lengths, 
albeit at the price of lower sensitivity to short lifetimes.  
Since we do not require the electron track to be reconstructed, our search results 
are also applicable for long-lived particle decaying into photons. 
% here comes detector &data description

The data in this analysis were recorded with the D0 
detector~\cite{d0det}, which comprises an inner tracker,
liquid-argon/uranium calorimeters, and a muon spectrometer.
The inner tracker is located in a 2~T superconducting solenoidal magnet and 
consists of silicon microstrip and scintillating-fiber trackers.
It provides measurements of charged particle tracks up to pseudorapidity \cite{d0_coord_sys}
of $|\eta|\approx3.0$.
The calorimeter system consists of a central section (CC) covering $|\eta|<1.2$ and
two endcap calorimeters extending the coverage to $|\eta|\approx 4$,
all housed in separate cryostats~\cite{d0cal}. The electromagnetic
section of the calorimeter has four longitudinal layers and
transverse segmentation of 0.1 $\times$ 0.1 in $\eta - \phi$ space
(where $\phi$ is the azimuthal angle), except in the third layer,
where it is 0.05 $\times$ 0.05. The central preshower (CPS) system is 
located between the solenoid and the CC calorimeter cryostat, covers 
$|\eta| \simleq 1.2,$ and provides measurement of EM shower position with a precision of about 1 mm. 
%The axes of EM showers are reconstructed
%by fitting straight lines to shower positions measured in the four longitudinal 
%calorimeter layers and the CPS (EM "pointing").
The data for this study were collected between 2002 and summer 2006 using single EM triggers . 
The integrated luminosity~\cite{d0lumi} of the sample is $1100\pm70$~pb$^{-1}$.

We select events with two EM clusters reconstructed in the central 
calorimeter with transverse momentum $p_T > 20$ GeV and $|\eta| < 1.1$, with the shower 
shape consistent with that expected of a photon. EM clusters are required
to be isolated in the calorimeter and tracker~\cite{gmsb_update}.
Both EM clusters are required to have a matched CPS cluster.
Only CPS clusters within a fixed  $\eta-\phi$ window are considered for matching, limiting
the electron or photon distance of closest approach ({\it DCA}) to the 
beam line at approximately 16 cm.
Jets are reconstructed using the iterative midpoint
cone algorithm~\cite{d0jets} with a cone size of 0.5.
The missing transverse energy is determined from the energy deposited 
in the calorimeter for $|\eta| < 4$ and is corrected for the EM and jet energy scales.

\begin{figure}
\includegraphics[width=8.6cm, height=6cm]{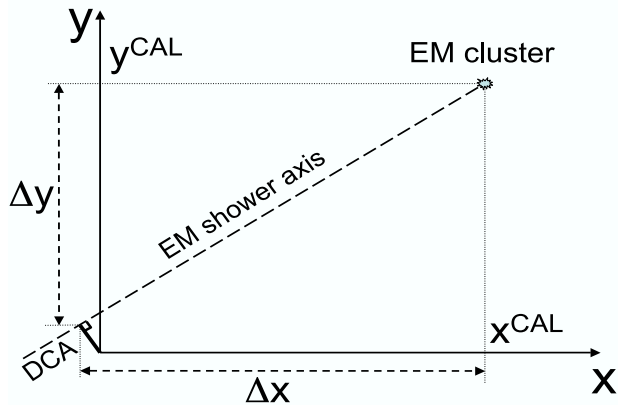}
\caption{\label{fig:sketch} Definition of the reconstructed EM particle trajectory. 
In the D0 coordinate system the equation of the trajectory is given by $\Delta x \cdot (y - y^{CAL}) = \Delta y \cdot (x - x^{CAL})$.
The distance from the beam line to the EM shower maximum $\sqrt{(x^{CAL})^2+(y^{CAL})^2}$ is typically around 90 cm. }
\end{figure}

The D0 EM pointing algorithm fits five shower position measurements 
(one in the CPS and four in the four EM layers of the central calorimeter) 
to a straight line which is assumed to be the EM object direction.
The electron trajectory for energies above 20 GeV, which are of interest to this analysis,
is very close to a straight line, which is defined by the energy-weighted 
EM cluster position ($x^{CAL}, y^{CAL}$) and 
the {\it DCA}. The {\it DCA} reconstruction accuracy is about 2 cm.
The common vertex position in the $xy$ plane for two EM objects is the
 intersection of the two lines associated to them and is given by a solution of the system of two linear equations (see Fig. \ref{fig:sketch} for definitions of the trajectory and quantities below):
\[ \left( \begin{array}{cc}
-\Delta y_1 & \Delta x_1  \\
-\Delta y_2 & \Delta x_2  \end{array} \right)
 \left( \begin{array}{c}
x  \\
y  \end{array} \right)
=
 \left( \begin{array}{c}
y^{CAL}_1\cdot \Delta x_1 - x^{CAL}_1 \cdot \Delta y_1  \\
y^{CAL}_2 \cdot \Delta x_2 - x^{CAL}_2 \cdot \Delta y_2   \end{array} \right).\]
The determinant of this system, $D$, is proportional to the sine of the 
opening angle $\theta_{12}$ between the EM objects. The vertex transverse position resolution is 
inversely proportional to the determinant.
Therefore, in the following we consider events with $|D|>4000~$cm$^2$ , which roughly 
corresponds to $\sin \theta_{12} > 0.5$, and use the variable
$R_S = \pm \sqrt{x^2+y^2} \cdot ( D/1000~$cm$^2)$, which, while related to the 
reconstructed vertex radius, also takes into account its uncertainty.
The sign of $R_S$ is given by the sign of the scalar product of the
$\vec{p}_T$ of the pair of EM objects with the vector pointing from the origin to the vertex location of the two EM particles.
To reduce the background we further require 
that at least one of the two EM objects has {\it DCA} $>2$ cm.

For vertices that originate from real particle decays, $R_S$
is positive, while its distribution for prompt electron or photon pairs 
is symmetrical around zero. The latter assumption was extensively checked 
with Monte Carlo (MC) simulation, a $Z\rightarrow e^+e^-$ data 
sample (both electrons 
in the $Z\rightarrow e^+e^-$ sample were required to have reconstructed tracks 
originating from the primary vertex), and a control sample of multi-jet 
events which has been selected exactly as the signal events except with 
an inverted tracker isolation requirement. Therefore, we estimate the 
background for positive values of $R_S$ by mirroring the negative part of the 
distribution.

\begin{figure}
\includegraphics[width=8.6cm]{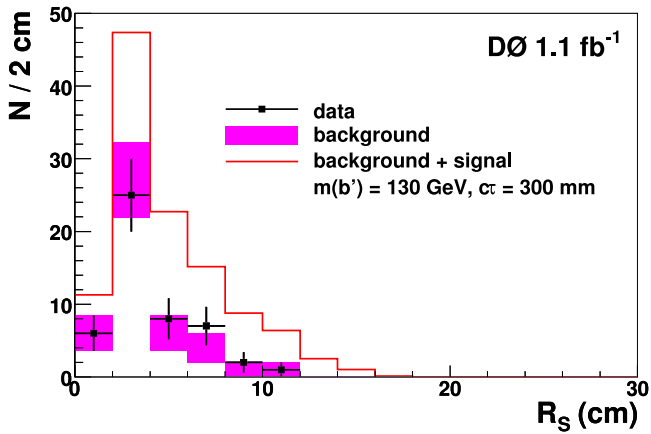}
\caption{\label{fig:rs75} Observed $R_S$ distribution for di-EM pairs with mass greater 
then 75 GeV (black points), expected distribution from prompt sources with its uncertainty (shaded rectangles) 
and the expected distribution in presence of $b^\prime$ quark with mass of 130 GeV and 
lifetime $c\tau=300$ mm (solid line).}
\end{figure}

The invariant mass $M$ of the two EM objects is corrected for the reconstructed vertex position, 
and the data are divided into three bins: $20<M<40,~40<M<75,$ and $M>75$ GeV. The last bin is used 
for searches for the fourth generation $b^\prime$. The corresponding observed $R_S$
distribution is shown in Fig.~\ref{fig:rs75}. All mass bins are used for a 
quasi model-independent search for long-lived particles. We also examine events with 
$\MET>30$ GeV and $M>20$ GeV. No excess of events with positive $R_S$ values is present 
in data (see Table~\ref{table:results}), so we proceed to set limits on new physics.

We use {\sc PYTHIA 6.319}~\cite{pythia} to generate events 
$p\bar{p} \rightarrow b^\prime \bar{b}^\prime \rightarrow ZbZb \rightarrow e^+e^- + X$.
{\sc PYTHIA} calculates production cross sections varying from 79.4 to 3.6 pb as the $b^\prime$ 
mass changes from 100 to 190 GeV. The events are then processed through the 
{\sc GEANT}-based~\cite{geant}  MC simulation, electronics and trigger simulation, 
and are reconstructed with the same reconstruction program as collision data. 
The expected $R_S$ distribution for a typical signal point is shown in Fig.~\ref{fig:rs75}. 
We use the efficiencies and acceptances obtained using this signal MC for 
the model-independent search as well. 
The significant jet activity in these 
events gives a conservative estimate of the efficiency for SUSY scenarios 
and should be adequate for hidden valley models~\cite{strassler_pcom}. 
In order to study different masses of hypothetical resonances in addition to the samples above we also generated samples of  $b^\prime \rightarrow vb$ for $v$ masses of 30 and 50 GeV.
We find that the efficiency and  acceptance for the MC events have no significant dependence 
on the masses of the $b^\prime$ and $v$. We set the $b^\prime$ mass to 150 GeV and 
vary its lifetime $c\tau$ between 2 and 7000 mm.

\begin{figure}
\includegraphics[width=8.6cm]{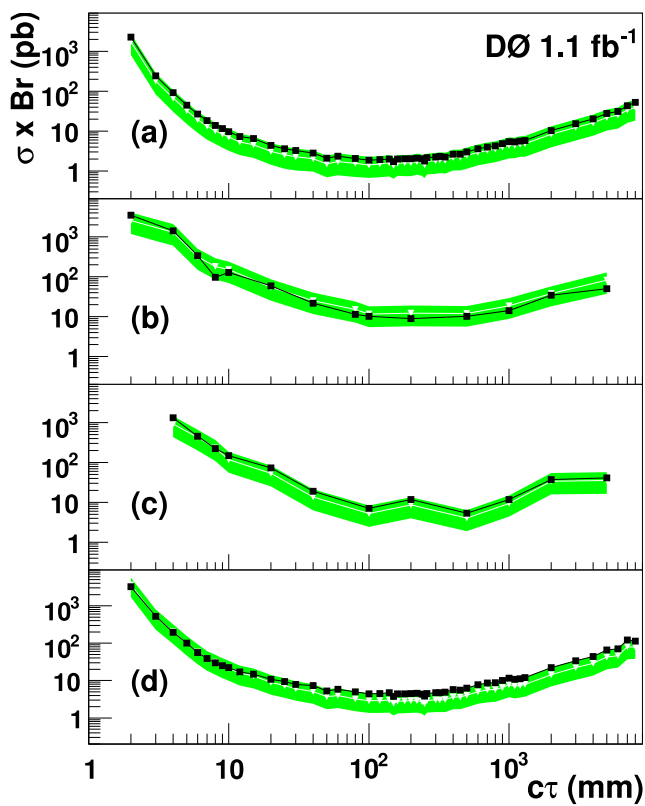}
\caption{\label{fig:limit} Expected (white triangles) and observed (black squares) 95\% C.L. upper limits 
on the cross section of a long-lived particle times the branching fraction of its decay to either a 
pair of electrons or photons for (a) $\MET>30$~GeV and $M >  20$~GeV, 
(b)  $20<M<40$~GeV, (c) $40<M<75$~GeV, and (d) $M>75$~GeV. 
All observed upper limits are within one standard deviation (shaded band) from the expected limits.}
\end{figure}

In Fig.~\ref{fig:limit} we display the limits on the production cross section 
of a long-lived particle times its branching fraction to decay into a pair of electrons. 
%Limits were set by fitting the $R_S$ distributions
%to a sum of signal and background predictions using the likelihood
%fitter~\cite{wade_fitter} that incorporates a log-likelihood
%ratio ($LLR$) test statistic method. The value of $CL_s$ is defined as
%$CL_s = CL_{s+b}/CL_b$, where $CL_{s+b}$ and $CL_b$ are the confidence levels
%for the signal plus background hypothesis and the background-only (null) hypothesis,
%respectively. These confidence levels are evaluated by integrating corresponding 
%$LLR$ distributions populated by simulating outcomes via Poisson statistics. 
%Systematic uncertainties are treated as uncertainties on the expected numbers of signal 
%and background events, not the outcomes of the limit calculations. This approach
%ensures that the uncertainties and their correlations are propagated to the
%outcome with their proper weights. 
  Limits were obtained from the $R_S$ distribution using the modified frequentist
  approach \cite{junk} as implemented in ~\cite{wade_fitter}. This method is based on a log-likelihood
  ratio ($LLR$) test statistic, and involves the calculation of confidence levels for the signal
  plus background and background-only (null) hypotheses (denoted by $CL_{s+b}$ and $CL_b$,
  respectively) by integrating the $LLR$ distributions resulting from simulated pseudo-experiments.
  The upper limit on the cross section at the 95\% C.L. is defined as the cross section value
  for which the ratio $CL_s = CL_{s+b}/CL_b = 0.05$.
The systematic uncertainties were taken to be flat as a
function of $R_S$. They include the uncertainty in electron or photon identification 
and triggering (15\%), 
uncertainty on Monte Carlo simulation (5\%), and uncertainty on luminosity (6.1\%).
At the $c\tau$ value of 100 mm we exclude at the 95\% C.L. the production cross section times branching fraction 
of long-lived particles that decay into a pair of electrons or photons above 1.9 pb, 10.2 pb, 7.1 pb, and 4.4 pb for $\MET > 30$~GeV and $M>20$~GeV, $20 < M < 40$~GeV, $40<M<75$~GeV, 
and $M>75$~GeV, respectively (see Fig. \ref{fig:limit}).

\begin{figure}
\includegraphics[width=8.6cm]{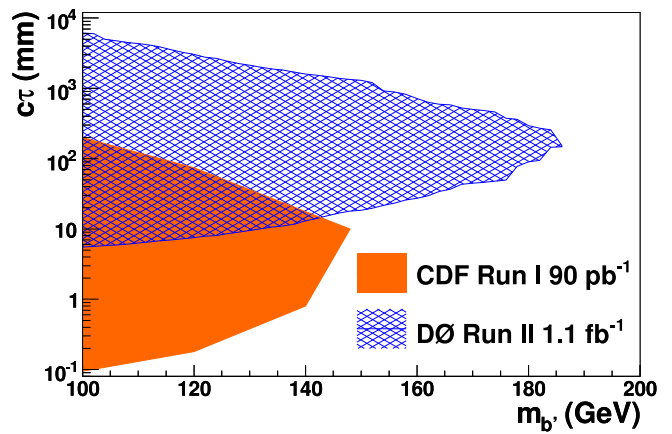}
\caption{\label{fig:bplimit} 95\% C.L. exclusion region of $b^\prime$ lifetime ($c\tau$) $vs.$~mass for CDF Run I \cite{cdf_bprime} and current D0 result.}
\end{figure}

Intersecting the cross section upper limits shown in Fig. 3d with the theoretical cross section 
of the production of the fourth generation $b^\prime$ quark \cite{pythia} we compute 
limits on its lifetime as a function of its mass assuming it decays only into $Zb$.
The limits are presented in Fig. \ref{fig:bplimit}, together with the exclusion region from the track-based CDF
search~\cite{cdf_bprime}. The two search methods are complementary to each other.

\begin{table}
\caption{\label{table:results}Observed number of events ($R_S>0$~cm) and estimated background ($R_S<0$~cm)
for different selections. }
\begin{ruledtabular}
\begin{tabular}{lcc}
Selection & $R_S>0$ & $R_S<0$\\
\hline
$20<M<40$ GeV & 38 & 47\\
$40<M<75$ GeV & 191 & 190\\
$M>75$ GeV & 49 & 45\\
$M>20$ GeV, $\MET>30$ GeV & 7 & 6\\ 
\end{tabular}
\end{ruledtabular}
\end{table}

To summarize, we have performed a search for long-lived particles decaying into electron 
or photon pairs using a new method that allowed us to explore previously unreachable 
portions of the parameter space. We find no evidence for such particles and present 
the results as model-independent limits on their production cross section and interpret 
them in the framework of a model with a long-lived $b^\prime$ quark \cite{bprime_theory}. 

We would like to thank Matt Strassler for many fruitful discussions.
% acknowledgement_paragraph_r2.tex                         5/23/08
%
We thank the staffs at Fermilab and collaborating institutions, 
and acknowledge support from the 
DOE and NSF (USA);
CEA and CNRS/IN2P3 (France);
FASI, Rosatom and RFBR (Russia);
CNPq, FAPERJ, FAPESP and FUNDUNESP (Brazil);
DAE and DST (India);
Colciencias (Colombia);
CONACyT (Mexico);
KRF and KOSEF (Korea);
CONICET and UBACyT (Argentina);
FOM (The Netherlands);
STFC (United Kingdom);
MSMT and GACR (Czech Republic);
CRC Program, CFI, NSERC and WestGrid Project (Canada);
BMBF and DFG (Germany);
SFI (Ireland);
The Swedish Research Council (Sweden);
CAS and CNSF (China);
and the
Alexander von Humboldt Foundation (Germany).
%
   % input acknowledgement


\begin{thebibliography}{99}
\bibitem{d0_coord_sys} the D0 detector utilizes a right-handed coordinate system with the $z$-axis 
pointing in the direction of the proton beam and the $y$-axis pointing
upwards. 
%
%The azimuthal angle $\phi$ of a point with coordinates
%($x$, $y$, $z$) is defined in the transverse, $xy$ plane with respect
%to $x$ axis. The longitudinal, $rz$ plane is defined by radius vector
%$r$ defined as $r \equiv \sqrt{x^2 + y^2 + z^2}$.
The azimuthal angle $\phi$ is defined in the $xy$ plane
measured from the $x$-axis. 
%The $rz$ plane is defined at constant
%azimuthal angle with coordinates $z$ and $r=\sqrt{x^2+y^2}$.
The pseudorapidity is defined as $\eta = -\ln[\tan(\theta/2)]$,
where $\theta = \arctan(\sqrt{x^2+y^2}/z)$.
\end{thebibliography}
\end{document}